\documentclass[12pt]{article}

\usepackage{bm}
\usepackage{amssymb}
\usepackage{enumerate}

\usepackage{graphicx} 

\begin{document}

\title{The SM  extensions with additional light scalar singlet, nonrenormalizable Yukawa interactions  and $(g-2)_{\mu}$}

\author{     S.N.Gninenko$^1$ and     N.V.~Krasnikov$^{1,2}$  
\\
$^{1}$ INR RAS, 117312 Moscow 
\\
$^{2}$ Joint Institute for Nuclear Research,141980 Dubna}




\date{\today}

\maketitle

\begin{abstract}

We consider   the SM extension  with additional light real singlet scalar, right-handed neutrino
and nonrenormalizable Yukawa interaction for the first two generations. 
We show that the proposed model  can explain the  observed $(g - 2)$ muon anomaly. Phenomenological 
consequenses  as flavour violating decays $\tau \rightarrow \mu\mu\mu, \mu \mu e, 
 \mu e e $ are briefly discussed. We also propose the $U_R(1)$ gauge generalization of 
the SM with complex scalar singlet and  nonzero 
right-handed charges for the first two generations.

\end{abstract}


\newpage

\section{Introduction}

The discovery of the neutrino oscillations \cite{Neutrino,pdg} means that at least two neutrino have nonzero masses. 
The minimal extension of the 
SM with nonzero neutrino masses  is the $\nu$MSM  \cite{nuMSM,nuMSM1}. In this model one adds to the SM 
three additional massive Majorana(right-handed)  fermions $\nu_{Ri}$, $i = 1,2,3$. 
Due to  seesaw mechanism \cite{nuMSM,seesaw} after the spontaneous 
$SU_L(2) \otimes U(1)$ electroweak symmetry breaking the neutrinos acquire masses $m_{\nu_i}= \frac{m^2_{Di}}{M_{Ri}}$. 
Here $m_{Di}$ are the Dirac neutrino masses and  $M_{Ri}$ are the  masses of the  $\nu_{Ri}$ neutrinos. 
The $\nu$MSM  has a candidate - 
the lightest Majorana neutrino  with a mass  $M_{\nu_{R}} \leq O(50)~KeV$ - for  dark matter.  
Besides, the model 
with light Majorana neutrino can solve the problem of the baryon asymmetry in our Universe \cite{nuMSM1}.

In this report which is based mainly on Refs.\cite{GniKr} we consider the extension of the $\nu$MSM 
with additional scalar field  and nonrenormalizable Yukawa interaction for the first two generations.  
We show that  the SM extension with 
additional light real singlet field can explain the  $(g - 2)$ muon anomaly. Phenomenological 
consequences of the proposed model as flavour violating decays $\tau \rightarrow \mu\mu\mu, 
\mu e e $ are brifly discussed. We also propose the $U_R(1)$ gauge generalization of 
the SM with complex scalar singlet and  nonzero 
righthanded charges for the first two generations.


\section{ The $\nu$MSM  extension with additional real scalar isosinglet and nonrenormalizable Yukawa interaction}

In this section  we consider the extension of the $\nu$MSM 
with additional scalar field  and nonrenormalizable Yukawa interaction for the first two generations \cite{GniKr}.  
The Lagrangian of the model has the form 
\begin{equation}
L_{tot} = L_{SM} + L_{Qd\phi} + L_{Qu\phi}    +  L_{Le\phi}  + L_{\phi} + L_{\nu_R} \,.
\end{equation}
Here 
\begin{equation}
L_{Qd\phi}  = -\frac{h_{Qd\phi,ik}}{M}\bar{Q}_{Li}\bar{H}\phi d_{Rk} 
+~H.c.~ \,,
\end{equation}
\begin{equation}
L_{Qu\phi} = 
-  \frac{h_{Qu\phi,ik}}{M}\bar{Q}_{Li}H\phi u_{Rk}
+~H.c.~ \,,
\end{equation}
\begin{equation}
L_{Le\phi} = -\frac{h_{Le\phi,ik}}{M}\bar{L}_{Li}\bar{H}\phi e_{Rk} +~H.c.~ \,,
\end{equation}
\begin{equation}
L_{\phi} = \frac{1}{2}\partial^{\mu}\phi\partial_{\mu}\phi -\frac{M^2_{\phi}\phi^2}{2} 
- \lambda\phi^4
\,,
\end{equation} 
\begin{equation}
L_{\nu_R} = i\bar{\nu}_{Rj} \hat{\partial} \nu_{Rj} -(\frac{M_{\nu_{Rj}}}{2}\nu_{Rj}\nu_{Rj} 
 + h_{Lij}\bar{L}_iH\nu_{Rj} + H.c.)  \,,
\end{equation}
where  $L_{SM}$ is the  SM Lagrangian, $\nu_{Rj}$ - are the Majorana neutrinos  and   
$L_1 = (\nu_{eL}, e_L)$, $L_2 = (\nu_{\mu L}, \mu_L)$, $L_3 = (\nu_{\tau L}, \tau_L)$,   $e_{R1} = e_R$, $e_{R2} = \mu_R$, $e_{R3} = \tau_R$,  
$Q_{L1} = (u_L, d_L)$, $Q_{L2} = (c_L, s_L)$, 
$Q_{L3} = (t_L, b_L)$, $d_{R1} = d_R$, $d_{R2} = s_R$, $d_{R3} = b_R$,  $u_{R1} = u_R$, $u_{R2} = c_R$, $u_{R3} = t_R$,
  $\bar{H} = (-(H^-)^*, (H^0)^*)$\footnote{Here $H = (H^0, H^-)$ is the SM Higgs doublet and $M$ is some high energy scale .}. 
The main peculiarity of the model (1-6) is the use of nonrenormalizable Yukawa interactions (2-4)\footnote{The nonrenormalizable 
Yukawa interactions have been considered in Refs.\cite{nonren1}.}. Here we consider the particular  case of 
the general model (1 - 6) with nonzero renormalizable Yukawa interaction   only for the 
third fermion generation.  We assume that  the masses of the first two light generations 
 arise due to nonrenormalizable interactions (2-4). We impose the discrete symmetry  
\begin{equation}
\phi \rightarrow   - \phi \,,
\end{equation}
\begin{equation}
e_{Rk} \rightarrow   -e_{Rk} ~~(k = 1,2)  \,,
\end{equation}
\begin{equation}
d_{Rk} \rightarrow   -d_{Rk} ~~(k = 1,2)  \,.
\end{equation}
The discrete symmetry (7 - 9) restricts the form of nonrenormalizable interactions (2-4), 
namely 
\begin{equation}
h_{Qd\phi,i3} = h_{Qu\phi,i3}=  0 \,,
\end{equation}
\begin{equation}
h_{Le\phi,i3} = 0 \,,
\end{equation}
As a consequence of the (7-9) the renormalizable SM Yukawa interaction with the first two 
quark and lepton generations vanishes  and the fermions of the first two generations  acquire 
masses only due to nonrenormalizable interactions (2-4).


We can consider 
the nonrenormalizable interactions (2 - 4) as some effective interactions arising from renormalizable 
interactions.  For instance, the  interaction (2) can be realized in renormalizable extension of the SM  with additional 
scalar field $\phi$ and new massive quark $SU(2)_L$ singlet fields $D_R$, $D_L$ with 
a mass $M_{D}$ and   $U(1)$ hypercharges 
$Y_{D_L} = Y_{D_R} = -\frac{1}{3}$. 
The interaction of new quark fields  $D_R$, $D_L$ with 
ordinary quarks and the neutral scalar field $\phi$ is
\begin{equation}
L_{qD \phi} = - c_i \bar{Q}_{Li}\bar{H}D_R  -   k_j\bar{D}_Ld_{Rj}\phi + H.c.\,.
\end{equation}
In the heavy $D$-quark mass limit $M_{D} \rightarrow \infty$ we obtain the effective interaction (2) with 
\begin{equation}
\frac{h_{Qd\phi,ij}}{M} = \frac{c_ik_j}{M_D} \,.
\end{equation}
Analogously we can consider nonrenormalizable interaction (4) as an effective interaction which arises in renormalizable 
extension of the SM with additional scalar field and new massive lepton $SU(2)_L$ singlet fields $E_R$, $E_L$ with 
$Y_{E_L} = Y_{E_R} = 1 $
The interaction of new lepton  fields  $E_R$, $E_L$ with 
ordinary quarks and the neutral scalar field $\phi$ is
\begin{equation}
L_{eE \phi} = - d_i \bar{L}_{Li}\bar{H}E_R  -   f_{j}\bar{E}_Le_{Rj}\phi + H.c.\,.
\end{equation}
In the heavy $E$-lepton  mass limit $M_{E} \rightarrow \infty$ we obtain the effective interaction (4) with 
\begin{equation}
\frac{h_{Le\phi,ij}}{M} = \frac{d_if_j}{M_E} \,.
\end{equation}

After the spontaneous $SU(2)_L\otimes U(1)$ electroweak symmetry breaking the Yukawa interaction  of the scalar 
field with charged leptons takes the form
\begin{equation}
L_{ll\phi } = -\bar{h}_{Le,ik}\bar{e}_{Li} \phi e_{Rk} + H.c \,,
\end{equation}  
where 
\begin{equation}
\bar{h}_{Le,ik} = h_{Le\phi,ik} \frac{<H>}{M} \,,
\end{equation}
$i = 1,2,3$, $k = 1,2$   and  $e_{L1} = e_L$, $e_{L2} = \mu_L$, $e_{L3} = \tau_L$, $<H> = 174~GeV$.
Nonzero vacuum expectation value for the real field $\phi$ generates   nonzero lepton masses for electrons and muons while the mass of $\tau$-lepton 
arises due to renormalizable Yukawa coupling. 
The lepton mass matrix and the Yukawa lepton 
$\phi^` = \phi~ - ~<\phi>$ interactions are
\begin{equation}
L_{ll} = -\bar{h}_{Le,ik}<\phi>\bar{e}_{Li}e_{Rk} - m_{\tau}\bar{e}_{L3}e_{R3}  + H.c \,,
\end{equation}
\begin{equation}
L_{ll\phi} = -\bar{h}_{Le,ik}\bar{e}_{Li} \phi^` e_{Rk} + H.c \,,
\end{equation}  
where $i  = 1,2,3$ and $k = 1,2$.
The mass terms (18) and  the interaction (19) have different flavour structure  that leads to the 
tree level flavour changing transitions  like $\tau \rightarrow \mu + \phi$, $\tau \rightarrow  e + \phi$ and as a consequence 
to the flavour violating decays like 
\begin{equation}
\tau^- \rightarrow \mu^- + \phi^* \rightarrow \mu^- \mu^+ \mu^- \,,
\end{equation}   
\begin{equation}
\tau^- \rightarrow \mu^- + \phi^* \rightarrow \mu^- e^+ e^- \,,
\end{equation}   
\begin{equation}
\tau \rightarrow e^- + \phi^* \rightarrow e^- \mu^+ \mu^- \,,
\end{equation}   
\begin{equation}
\tau \rightarrow e^-  + \phi^* \rightarrow e^- e^+ e^- \,.
\end{equation}   
At present state of art  we can't predict the value of flavour violating Yukawa couplings $\bar{h}_{Le31}, \bar{h}_{Le32}$.
  

The interaction (19) leads, in particular, to 
the additional one loop contribution to muon magnetic moment due to $\phi^`$ 
scalar exchange, namely \cite{mureview}
\begin{equation}
\Delta a_{\mu} = \frac{h^2_{Le,22}}{8\pi^2}\frac{m^2_{\mu}}{M^2_{\phi}}
\int^1_0\frac{x^2(2-x)}{(1-x)(1-\lambda^2x) + \lambda^2 x} \,,
\end{equation}
where $\lambda =  \frac{m_{\mu}}{m_{\phi}}$. 
In the limit $M_{\phi} >> m_{\mu}$ 
\begin{equation}
\Delta a_{\mu} = \frac{1}{4\pi^2}\frac{m^2_{\mu}}{M^2_{\phi}} 
\bar{h}^2_{Le,22}[ln(\frac{M_{\phi}}{m_{\mu}}) 
- \frac{7}{12}] \,.
\end{equation} 
The precise measurement of the anomalous magnetic
moment of the positive muon from the
Brookhaven AGS experiment \cite{g-2} gives a result which
is $3.6 \sigma$ higher than the Standard Model (SM) prediction 
\begin{equation}
a_{\mu}^{exp} - a_{\mu}^{SM} = (288 \pm 80) \cdot 10^{-11} \,,
\end{equation} 
where $a_{\mu} \equiv \frac{g_{\mu} -2}{2}$. 
  Using the formulae (25, 26) we find that for $m_{\Phi} = (100, 10, 1, 0.5)~GeV$ the muon 
$g-2$ anomaly can be explained for

\begin{equation}
\bar{h}^2_{Le,22} = (1.6  \pm 0.5) \cdot 10^{-2}\,  ~~for    ~~~~m_{\phi} = 100~GeV \,,
\end{equation}
\begin{equation}
\bar{h}^2_{Le,22} = (2.6 \pm 0.8) \cdot 10^{-4}\,  ~~for    ~~~~m_{\phi} = 10~GeV \,,
\end{equation}
\begin{equation}
\bar{h}^2_{Le,22} = (6.2  \pm 1.9) \cdot10^{-6} \, ~~for  ~~~~m_{\phi} = 1~GeV \,.
\end{equation}
\begin{equation}
\bar{h}^2_{Le,22} = (2.6  \pm 0.8) \cdot10^{-6} \, ~~for  ~~~~m_{\phi} = 0.5~GeV \,.
\end{equation}
For the opposite limit $m_{\mu} \gg m_{\phi}$
\begin{equation}
\Delta a_{\mu} = \frac{3h_{Le,22}^2}{16\pi^2}
\end{equation}
and as a consequence of (26, 31) we find that
\begin{equation}
\bar{h}^2_{Le,22} = (1.5 \pm 0.5)\cdot 10^{-7} \,.
\end{equation}
As in the SM the Yukawa couplings $\bar{h}_{Le.ii}$ are proportional to the lepton masses. 
It means that  the interaction of the $\phi $ scalar with electrons is weaker than 
the interaction of the $\phi $ scalar with muons by factor $m_{\mu}/m_{e} \approx 200$
and the contribution of the $\phi$ scalar to the electron magnetic moment is 
suppressed at least by factor $(m_{e}/m_{\mu})^2$ in comparison with  the muon magnetic moment 
even for superlight $m_{\phi} \ll m_{e}$ scalar. 
For instance, for $m_{\phi} = 1~GeV$ the contribution of the $\phi$ scalar to the electron 
magnetic moment is 
\begin{equation}
(\Delta a_e)_{\phi} =  (0.16 \pm 0.05)\cdot 10^{-17}
\end{equation}
that is much smaller  the  bound  from $a_e$ \cite{elmag}
\begin{equation}
   \Delta a_e =  a^{exp}_e - a^{SM}_e = (-1.06 \pm 0.82)\cdot10^{-12} 
\end{equation}
Due to the suppression factor $\frac{m_e}{m_{\mu}}$ for electon Yukawa coupling in comparison  with 
muon Yukawa coupling the search for light $\phi$ scalar in 
electron fixed target experiments 
or $e^+e^-$ experiments is very problematic but not hopeless.\footnote{Roughly speaking we have to 
improve the  discovery potential by 3-4 orders of magnitude.} The search for very light $\phi$ scalar 
in $\pi \rightarrow (\phi \rightarrow e^+e^-) \gamma$ decay is possible but again we have 
additional suppression factor $(\frac{m_u -m_d}{m_{\mu}})^2 \sim O(10^{-2})$. 
 Light scalar particle $\phi$ with a mass $m_{\phi} \lesssim 1~GeV$    decaying into muon pair 
can be searched for at CERN SPS secondary muon beam in full analogy with the search for new 
light vector boson $Z^`$ \cite{GKM}.

\section{The $U(1)$ gauge generalization of the model with real scalar field}
Here we outline one of possible generalizations of the model (1-6). 
In the proposed generalization instead of real scalar $\phi$ we use 
complex scalar $\Phi$ and new abelian gauge groop $U_R(1)$\footnote{In Refs. \cite{Lmu}
new light vector boson  interacting with the  $L_{\mu} - L_{\tau}$ current has been proposed  for 
$(g-2)_{\mu}$ anomaly explanation, see also Ref.\cite{Posp} where the model with new light 
gauge boson interacting with the SM electromagnetic current has been proposed for the $(g-2)_{\mu}$ anomaly 
explanation and Ref.\cite{Lee} where the interaction of light gauge boson with $(B - L) + xY$ current has been considered.} 
 with nonzero charges for 
right handed fermions of  the first and second generations, namely
\begin{equation}
Q_X(u_R) = Q_X(c_R) = Q_X(\nu_{e R}) = Q_X(\nu_{\mu R}) = - Q_X(d_R) = -Q_X(s_R) = - Q_X(e_R) =
 -Q_{X}(\mu_R) \,.
\end{equation}
The nonrenormalizable Yukawa interactions of the first and second 
generation fermions in full analogy with the (2-4) interactions take the form 
\begin{equation}
L_{Qd\Phi}  = -\frac{h_{Qd\Phi,ik}}{M}\bar{Q}_{Li}\bar{H}\Phi d_{Rk} 
+~H.c.~ \,,
\end{equation}
\begin{equation}
L_{Qu\Phi} = 
-  \frac{h_{Qu\Phi,ik}}{M}\bar{Q}_{Li}H\Phi^{*} u_{Rk}
+~H.c.~ \,,
\end{equation}
\begin{equation}
L_{Le\Phi} = -\frac{h_{Le\Phi,ik}}{M}\bar{L}_{Li}\bar{H}\Phi e_{Rk} +~H.c.~ \,,
\end{equation}
\begin{equation}
L_{L\nu\Phi} = -\frac{h_{L\nu\Phi,ik}}{M}\bar{L}_{Li}\bar{H}\Phi^{*} \nu_{Rk} +~H.c.~ \,,
\end{equation}
Note that proposed model is free from $\gamma_5$-anomalies and we can consider the
 origin  of the $U_R(1)$ gauge group as a result of the gauge symetry breaking 
$SU_L(2) \otimes SU_R(2) \otimes U(1) \rightarrow   SU_L(2)\otimes U_R(1) \otimes U(1) $. 
We assume that in the considered model $<\Phi> \neq 0 $ that  leads to 
nonzero  $X$ gauge boson mass  and nonzero fermion masses for the first and second generations. 
In the unitaire gauge $\Phi = \phi~ + ~<\Phi>$, where $\phi = \phi^*$ is real scalar field as in 
previous section plus we have massive vector boson $X$. 
So in this model after $U_R(1)$ gauge symmetry breaking in addition to   the $\nu$MSM spectrum 
we have both  scalar and vector  particles.
The one loop contribution to the 
anomalous muon(electron) magnetic moment  due to the $\phi$ and $X$ exchanges is 
\begin{equation}
\Delta a_{\mu} = \Delta a_{\mu}(\phi) + \Delta a_{\mu}(X) \,,
\end{equation}
where the $\Delta a_{\mu}(\phi)$ contribution is given by the formulae (24,25,31) and the vector $X$ boson contribution 
for $(V+A)$ right-handed coupling\footnote{The interaction Lagrangian of the vector $X$ field with 
muons is $L_{X\bar{\mu}\mu} = g_{X} X^{\nu}\bar{\mu}\gamma_{\nu}(1 + \gamma_5)\mu $.}
with fermions is \cite{mureview}
\begin{equation}
\Delta a_{\mu} = \frac{g^2_{X}}{8\pi^2}\frac{m^2_{\mu}}{M^2_{X_{\mu}}}
\int^1_0\frac{2 x^2(2-x) + 2x(1-x)(x-4) -4\lambda^2 x^3}{(1-x)(1-\lambda^2x) + \lambda^2 x} \,,
\end{equation}
\begin{equation}
\Delta a_{\mu} = -  \frac{g^2_{X}}{3\pi^2}\frac{m^2_{\mu}}{M^2_{X}}~~  ~~ (for ~M_{X}~ \gg ~m_{\mu}) \,.
\end{equation}
The   $X$ boson                            
contribution (41)  to the $(g - 2)$ is negative. For instance, for $\alpha_X = \frac{g^2_X}{4\pi} =  10^{-8}$
and $M_X = 500~MeV$  the $X$-boson contribution to $\Delta a_{\mu}$ is $\Delta a_{\mu}(X) = -1.8\cdot10^{-10}$.
The positive contribution due to $\phi$ boson exchange is positive and cancels the negative contribution from the $X$ boson exchange. 
The most perspective experiments for the search  for light vector $X$-boson  with the electron coupling 
constant $\alpha_{X}  = O(10^{-8})$ are the                                                                                      
electron fixed target experiments 
or $e^+e^-$ experiments. The $X$ boson can decay into electron-positron or muon-antimuon pairs, also invisible decays of 
the $X$ boson into light sterile neutrino are possible.  The    experiment NA64 \cite{P348} at CERN will be able to search for both invisible and visible 
$X$ boson decay modes with the  $\alpha_X \geq O(10^{-12})$ \cite{NA64}.


\section{Conclusion}

The $\nu$MSM  
with additional scalar field and nonrenormalizable interaction for the first two generations 
can explain the observed muon $(g-2)$ anomaly. The model predicts the existence of flavour violating 
quark and lepton decays like  $\tau \rightarrow \mu\mu\mu, \mu \mu e, 
 \mu e e $. Besides the $U(1)$ gauge generalization of the model with 
real isosinglet scalar field is also able to explain muon $(g-2)$ anomaly.

We are indebted to  colleagues  from INR theoretical department for discussions and 
comments.

\end{document}